\documentclass[letterpaper, 10 pt, conference]{ieeeconf}  

\IEEEoverridecommandlockouts                              

\usepackage{amsmath,amsfonts}
\usepackage{array}
\usepackage{textcomp}
\usepackage{stfloats}
\usepackage{hyperref}
\usepackage{url}
\usepackage{verbatim}
\usepackage{comment}
\usepackage{cite}
\usepackage{float} 
\usepackage{placeins} 
\usepackage{amsmath,amssymb,amsfonts}
\usepackage{graphicx}
\usepackage{textcomp}
\usepackage{xcolor}
\usepackage{caption}
\usepackage{cite}
\usepackage{multirow}
\usepackage{url}
\usepackage{graphicx}
\usepackage{textcomp}
\usepackage{caption}
\usepackage{multicol}
\usepackage{xcolor}
\usepackage{subcaption}
\usepackage{array}
\usepackage{float}
\usepackage{algorithm}      
\usepackage{algpseudocode}  
\usepackage{graphicx}
\usepackage{textcomp}
\usepackage{scalerel}
\usepackage{cite}
\usepackage{amsmath,amssymb,amsfonts}
\usepackage{graphicx}
\usepackage{textcomp}
\usepackage{caption}
\usepackage{multicol}
\usepackage{xcolor}
\usepackage{esdiff}
\usepackage{etoolbox}
\usepackage{cases}
\usepackage{mathtools}                                   
\usepackage{balance} 
\usepackage{graphicx} 
\usepackage{float} 
\usepackage{lipsum} 
\usepackage{graphicx}
\usepackage{textcomp}
\usepackage{xcolor}
\usepackage{cite}
\usepackage{amsmath,amssymb,amsfonts}
\usepackage{graphicx}
\usepackage{textcomp}
\usepackage{scalerel}
\usepackage{multicol}
\usepackage{etoolbox}
\usepackage{cases}
\usepackage{mathtools}
\usepackage{comment}
\usepackage{tikz}
\usetikzlibrary{arrows.meta,positioning,calc}

\makeatletter
\newcommand*{\rom}[1]{\expandafter\@slowromancap\romannumeral #1@}
\makeatother

\usepackage[none]{hyphenat} 
\usepackage{microtype}      
\title{\LARGE \bf
	A Survey on Applications of Quantum Computing for Unit Commitment
}

\author{
Milad Hasanzadeh$^{1}$, Ali Rajabi$^{1}$, and Amin Kargarian$^{1}$%
\thanks{This work was supported by the National Science Foundation under Grant ECCS-1944752 and Grant ECCS-2312086.}%
\thanks{$^{1}$All authors (Amin Kargarian is corresponding author) are with the Department of Electrical Engineering, Louisiana State University, Louisiana, USA.
        {\tt\small mhasa42@lsu.edu, Ali.Rajabi@lsu.edu, kargarian@lsu.edu}}%
}

\begin{document}

	\maketitle
	\thispagestyle{empty}
	\pagestyle{empty}

\begin{abstract}
Unit Commitment (UC) is a core optimization problem in power system operation and electricity market scheduling. It determines the optimal on/off status and dispatch of generating units while satisfying system, operational, and market constraints. Traditionally, UC has been solved using mixed-integer programming, dynamic programming, or metaheuristic methods, all of which face scalability challenges as systems grow in size and uncertainty. Recent advances in quantum computing—spanning quantum annealing, variational algorithms, and hybrid quantum–classical optimization—have opened new opportunities to accelerate UC solution processes by exploiting quantum parallelism and entanglement. This paper presents a comprehensive survey of existing research on the applications of quantum computing for solving the UC problem. The reviewed works are categorized based on the employed quantum paradigms, including annealing-based, variational hybrid, quantum machine learning, and quantum-inspired methods. Key modeling strategies, hardware implementations, and computational trade-offs are discussed, highlighting the current progress, limitations, and potential future directions for large-scale quantum-enabled UC.
\end{abstract}


\section{Introduction}
\label{sec:intro}

The past decade has seen rapid advances in quantum computing, sparking growing interest in its application to large-scale engineering optimization. Among these, the unit commitment (UC) problem occupies a central role in power system operation and electricity market scheduling. UC determines the on/off status and generation levels of thermal and renewable units over a scheduling horizon while satisfying demand balance, ramping, reserve, and network constraints. Because it couples binary and continuous decisions under nonlinear cost characteristics, UC remains one of the most computationally demanding optimization problems in the power and energy domain~\cite{wood2013power,hasanzadeh2025admm,hasanzadeh2025admm1}.

Classical UC formulations have long relied on mathematical programming and heuristic approaches. Mixed-integer linear and nonlinear programming techniques offer deterministic optimality but suffer from exponential growth in computation with system size and uncertainty~\cite{javadi2025learning,hasanzadeh2026all}. Dynamic programming and priority-list methods provide faster approximations yet become intractable for realistic multi-area or stochastic variants. Metaheuristic algorithms—such as genetic algorithms (GA), particle swarm optimization (PSO), simulated annealing (SA), and differential evolution (DE)—have improved flexibility and robustness~\cite{muralikrishnan2020comprehensive}. Still, they lack guarantees of global optimality and may require extensive tuning. Even with modern solvers, UC, large-scale security-constrained UC (SCUC), and stochastic remain NP-hard. 

Quantum computing offers a fundamentally different computational framework. By leveraging the principles of superposition and entanglement, quantum processors can represent and evaluate an exponentially large number of candidate solutions simultaneously~\cite{hasanzadeh2025finite}. Many discrete optimization problems, including UC, can be reformulated as quadratic unconstrained binary optimization (QUBO) models or equivalently as Ising Hamiltonians~\cite{hasanzadeh2026distributed}. This QUBO–Ising mapping enables direct execution of UC formulations on quantum hardware through energy-minimization processes. Early studies have demonstrated the feasibility of using quantum annealing on D-Wave platforms~\cite{morstyn2022annealing} and gate-based variational algorithms such as the quantum approximate optimization algorithm (QAOA) and the variational quantum eigensolver (VQE)~\cite{mahroo2023learning,hasanzadeh2025distributed}. These works highlight the potential of hybrid quantum–classical schemes to accelerate scheduling.

Despite growing attention, the literature on quantum computing for power system optimization remains fragmented across different algorithmic and hardware perspectives. Some efforts emphasize hardware-specific annealing implementations~\cite{christeson2024quantum,ling2025hybrid}, others focus on hybrid or distributed gate-based formulations that embed quantum subroutines within classical coordination methods such as alternating direction method of multipliers (ADMM) \cite{mahroo2023learning} or Benders decomposition~\cite{gao2022hybrid,nikmehr2022quantum}, and a smaller number investigate learning-driven frameworks using quantum neural networks (QNNs) or quantum reinforcement learning (QRL)~\cite{liu2025exact,wei2024quantum}. What is currently missing is a unified survey that synthesizes these developments, contrasts their modeling assumptions, and identifies the practical pathways toward scalable quantum-enabled UC.

This paper provides such a synthesis. It systematically reviews quantum computing applications to the UC problem. This survey paper organizes existing research into four methodological groups: (i) quantum annealing–based formulations, (ii) variational and hybrid quantum–classical optimization frameworks, (iii) quantum machine-learning–assisted scheduling, and (iv) quantum-inspired metaheuristics. The goal is to establish a cohesive understanding of how quantum paradigms have been adapted for UC, what computational trade-offs they introduce, and where open research opportunities remain. 

\section{Overview of Quantum Computing}
\label{sec:qc_overview}

Quantum computing leverages the principles of superposition and entanglement to explore vast decision spaces in parallel, offering new opportunities for accelerating combinatorial optimization in power systems. Unlike classical bits, qubits can represent a superposition of $0$ and $1$, allowing simultaneous evaluation of multiple states. This parallelism, combined with interference-based search, makes quantum algorithms well suited for problems such as UC and optimal power flow (OPF), where binary and nonlinear constraints lead to high computational complexity.

Many power system optimization problems can be reformulated as QUBO or, equivalently, as Ising Hamiltonians. Objective functions and constraints are encoded through quadratic cost and penalty terms, and the resulting Hamiltonian’s ground state represents the optimal solution. This QUBO–Ising mapping provides a unified interface between classical optimization models and quantum hardware.

Two main quantum paradigms have been applied to power system optimization. Quantum annealing uses adiabatic evolution to minimize an Ising Hamiltonian and has been implemented on platforms such as D-Wave to solve discrete scheduling problems. Gate-based variational algorithms, including the QAOA and the VQE, combine parameterized quantum circuits with classical feedback loops, enabling hybrid optimization compatible with noisy intermediate-scale quantum (NISQ) devices.

Recent research also explores distributed and hybrid quantum–classical methods—such as Quantum ADMM and Quantum Benders Decomposition—to handle large-scale systems through decomposed QUBO subproblems. Moreover, emerging quantum machine learning (QML) frameworks, including QNNs and QRL, extend quantum computing to adaptive and data-driven scheduling.

Current quantum processors from D-Wave, IBM, IonQ, and others already enable small-scale demonstrations of these techniques. Although constrained by qubit count, noise, and connectivity, such devices provide a foundation for near-term hybrid optimization. These developments (see Fig. \ref{fig:qc_overview_final}) establish the basis for Section~\ref{sec:review_qc_uc}, which reviews how quantum computing paradigms are specifically applied to UC.

\begin{figure}[!t]
\centering
\begin{tikzpicture}[
  font=\footnotesize,
  node distance=5mm and 7mm,
  box/.style={
    draw, rounded corners=2pt, align=center, inner sep=3pt,
    minimum width=50mm, thick, text=black
  },
  head/.style={box, very thick, fill=#1!25, draw=#1!70!black},
  sub/.style={box, fill=#1!18, draw=#1!60!black},
  link/.style={-{Latex[length=2mm,width=1.2mm]}, thick, draw=#1!70!black},
]

\node[head=blue] (pr) {Quantum Principles};
\node[head=cyan, below=4mm of pr] (qb)
  {Optimization Bridge: QUBO $\leftrightarrow$ Ising};
\draw[link=blue] (pr) -- (qb);

\node[sub=green!60!black, below left=8mm and 24mm of qb.south, anchor=north,
      minimum width=35mm, minimum height=8mm, inner sep=2pt]
    (qa) {Annealing: Adiabatic / Ising};

\node[sub=teal, below right=8mm and 24mm of qb.south, anchor=north,
      minimum width=35mm, minimum height=8mm, inner sep=2pt]
    (va) {Variational: QAOA, VQE};

\draw[link=cyan] (qb.south) -- ++(0,-1mm) -| (qa.north);
\draw[link=cyan] (qb.south) -- ++(0,-1mm) -| (va.north);

\path let \p1 = (qa.south), \p2 = (va.south) in
      coordinate (midSV) at ($(\p1)!.5!(\p2)$);
\coordinate (hub) at ($(midSV)+(0,-2mm)$);

\node[sub=violet, below=6mm of midSV, minimum width=62mm]
  (ext) {Hybrid / Distributed \& QML: QADMM, QGBD, QNN, QRL};

\draw[link=green!60!black] (qa.south) |- (hub);
\draw[link=teal] (va.south) |- (hub);
\draw[link=violet] (hub) -- (ext.north);

\node[head=orange!90!black, below=15mm of hub, minimum width=58mm]
  (apps) {Power System Applications: UC, OPF, SCUC};
\draw[link=orange!80!black] (ext) -- (apps);

\end{tikzpicture}
\caption{Overview of quantum principles to power systems.}
\label{fig:qc_overview_final}
\vspace{-9pt}
\end{figure}

\section{Quantum Computing in UC} \label{sec:review_qc_uc}

We categorize prior UC studies into four groups, each reflecting a distinct quantum paradigm applied to the problem: annealing methods that solve QUBO-encoded UC via Ising minimization, variational/hybrid algorithms suited for NISQ hardware, QML approaches that frame UC as a learning task, and quantum-inspired metaheuristics that mimic quantum search classically. This separation aligns with the natural methodological boundaries in the UC literature, enabling a clear and fair comparison of techniques.

\vspace{-3pt}
\subsection{Quantum Annealing--Based UC}
\label{subsec:annealing_uc}
\vspace{-3pt}
Quantum annealing represents one of the earliest paradigms applied to UC, focusing on direct QUBO or Ising formulations executable on hardware such as D-Wave. The following studies highlight various annealing and search-based approaches that address the complexity and scalability of UC.

In \cite{zheng2024fast}, a Grover-inspired quantum search algorithm is proposed for UC, using amplitude amplification to locate near-optimal on/off schedules. An oracle encodes feasible solutions based on generation cost and constraints, and iterative amplification enhances the probability of selecting optimal commitments. A classical refinement step adjusts dispatch for feasibility, providing a quantum-accelerated alternative to exhaustive or annealing-based search.

In \cite{hong2025qubit}, a qubit-efficient quantum annealing method is introduced for stochastic UC using a two-stage stochastic UC framework. The Powell–Hestenes–Rockafellar augmented Lagrangian approach replaces slack variables with dual penalties, yielding a compact QUBO. A quantum ADMM decomposition further divides the master problem into smaller subproblems that can be solved on annealing hardware, ensuring consistency through multiplier updates. This design reduces embedding complexity and scales with uncertainty scenarios.

In \cite{christeson2024quantum}, a quantum-compatible UC formulation is proposed using logarithmic discretization to reduce qubit usage in QUBO models. Generation levels are encoded with logarithmic intervals instead of linear steps, capturing wide ranges with fewer binaries. The resulting compact QUBO integrates commitment and discretized generation variables with quadratic penalties for balance and limits, yielding a hardware-efficient and scalable annealing formulation.

In \cite{ling2025hybrid}, a hybrid quantum annealing decomposition framework is introduced for scalable UC optimization. The system-level UC is divided into local QUBO subproblems, which are solved by quantum annealing, with coupling constraints managed through classical coordination. A variable-reduction strategy minimizes auxiliary variables and qubit use, enabling efficient hardware embedding while preserving feasibility.

In \cite{muller2025quantum}, a quantum annealing framework is developed for renewable-integrated UC by formulating binary on/off and renewable scheduling decisions as a QUBO. The objective combines cost and curtailment penalties, with system constraints enforced via quadratic terms. Mapped to an Ising Hamiltonian and solved on a quantum annealer, the approach yields feasible, near-optimal schedules with faster convergence than classical heuristics.

In \cite{morstyn2022annealing}, a quantum annealing approach is applied to a combinatorial OPF problem by casting discrete scheduling and network decisions into a QUBO for Ising-based optimization. Binary activation states and quadratic penalty terms capture operational constraints, while a hybrid quantum–classical loop refines solutions post-annealing. Though centered on OPF, the method generalizes to UC, showcasing the viability of annealing solvers for discrete power system scheduling.

\subsection{Variational and Hybrid Quantum--Classical Frameworks}
\label{subsec:variational_uc}
Building upon annealing formulations, variational and hybrid quantum–classical methods exploit gate-based architectures and decomposition schemes to solve UC more flexibly. The following works integrate QAOA, VQE, or distributed ADMM structures within hybrid optimization frameworks.

In \cite{nikmehr2022quantum}, the UC problem is formulated as a QUBO and solved via QAOA within a hybrid quantum–classical loop, where quantum circuits evaluate the Hamiltonian and a classical optimizer updates variational parameters. To enhance scalability, QAOA is embedded in a distributed ADMM framework (quantum-ADMM), enabling each microgrid to solve local UC subproblems quantumly while coordinating through dual updates, illustrating distributed variational UC optimization.

In \cite{halffmann2022quantum}, the UC problem is reformulated as a QUBO for direct execution on quantum hardware, encoding on/off states, up/down times, and demand constraints in a compact quadratic form. The model minimizes qubit use by eliminating slack variables and redundant terms, with the ground state of the corresponding Ising Hamiltonian representing a feasible UC schedule. This hardware-oriented formulation highlights the importance of encoding efficiency for NISQs.

In \cite{koretsky2021adapting}, the UC problem is cast as a binary QUBO and mapped to an Ising Hamiltonian for solution via QAOA. Unit commitments and generation limits are encoded as quadratic cost and penalty terms, while the QAOA circuit alternates between cost and mixer operators with parameters $(\gamma,\beta)$ optimized classically. Implemented in Qiskit, this early study shows how gate-based QAOA can approximate feasible UC schedules within a hybrid quantum–classical framework.

In \cite{nikmehr2022quantum2}, a distributed quantum-enabled UC framework is proposed, decomposing the scheduling task into regional QUBO subproblems solved via hybrid quantum–classical solvers. Regional coordination is maintained through iterative multiplier updates enforcing balance and coupling constraints. This architecture reduces qubit demand, enables parallel execution, and demonstrates scalable hybrid quantum scheduling across microgrids or market regions.

In \cite{fu2025quantum}, a quantum-embedded robust optimization framework is proposed for resilience-constrained UC, combining QAOA with a classical robust loop. The UC is reformulated as a QUBO with additional penalties for resilience constraints. The quantum layer solves the commitment subproblem and the classical layer updates penalties and worst-case scenarios. This hybrid design enables resilience-aware UC by using quantum subroutines to explore feasible configurations under uncertainty.

In \cite{feng2022novel}, a quantum surrogate Lagrangian relaxation (QSLR) framework is introduced for UC, embedding quantum subroutines within a classical decomposition scheme. The SLR method relaxes coupling constraints like load balance and reserves, producing binary subproblems expressed as QUBOs solved via quantum optimizers, while dual multipliers are updated classically. This hybrid design combines quantum efficiency in local search with classical coordination for global convergence, outlining a decomposition-based quantum UC.

In \cite{salgado2024hybrid}, a hybrid classical–quantum framework is developed for constrained UC with ramping, up/down, and reserve limits. The binary subproblem is cast as a QUBO and solved via warm-start QAOA initialized from a classical relaxation, while a classical quadratic solver handles the continuous dispatch. Iterative coordination between the two layers yields feasible and scalable UC solutions under limited resources.

In \cite{aboumrad2025new}, a multi-stage hybrid quantum–classical algorithm is proposed for UC, combining a variational quantum optimizer with a classical dispatch refinement step. The VQA generates candidate commitment vectors, while a classical SLSQP solver refines power outputs and checks feasibility. Iterative information exchange accelerates convergence, and tests on IonQ hardware demonstrate the practicality of VQA-based hybrid frameworks for UC scheduling.

In \cite{yang2023quantum}, a quantum-aided hybrid algorithm is presented for UC, combining a quantum variational solver for binary commitments with a classical optimizer for dispatch. The quantum layer produces candidate schedules, the classical layer refines power outputs and feasibility, and feedback updates the quantum parameters. This iterative coordination accelerates binary-space exploration and illustrates the synergy between quantum and classical solvers in large-scale UC scheduling.

In \cite{yang2025scalable}, a fully distributed quantum ADMM (QADMM) framework is introduced for large-scale UC. The UC is decomposed into binary QUBO and continuous subproblems solved by quantum and classical optimizers, respectively. A consensus-based coordination enforces global constraints through dual updates among agents. By distributing QUBO tasks and minimizing inter-agent communication, the method scales efficiently while reducing regional qubit demands.

In \cite{gao2022hybrid}, a quantum generalized Benders decomposition (QGBD) framework is proposed for UC, combining a quantum-solved binary master QUBO with classical continuous subproblems that generate Benders cuts. Distributed variants assign local masters to regional agents coordinated via consensus, reducing QUBO size and qubit demand. This hybrid design adapts classical decomposition for scalable quantum execution in UC optimization.

In \cite{magar2024dc}, a hybrid quantum–classical ADMM framework is developed for integrated UC and DC-OPF optimization. The binary commitment subproblem is formulated as a QUBO and solved via a quantum optimizer, while the DC-OPF subproblem is handled classically. ADMM-based dual exchanges enforce nodal balance and line limits, demonstrating how quantum solvers can be embedded into classical network optimization for scalable joint scheduling.

In \cite{liu2025exact}, the partially connected quantum neural network (PCQNN) framework employs quantum neural networks as variational circuits for UC, where quantum models generate candidate schedules and a classical optimizer updates parameters using cost and feasibility feedback. This hybrid learning–optimization design links variational optimization with quantum machine learning, enabling adaptive and hardware-efficient UC scheduling.

In \cite{wei2024quantum}, a two-stage QRL-based UC framework integrates a quantum reinforcement learning module with classical dispatch and adjustment submodules. The iterative interaction between quantum and classical layers enables large-scale scheduling under uncertainty, bridging quantum optimization and quantum machine learning in UC applications.

In \cite{zhou2025problem}, a problem-structure-informed QAOA is proposed for large-scale UC under limited qubit resources. By exploiting network topology and constraint sparsity, the method reduces circuit depth and coupling complexity. The UC is partitioned into regional QUBO subproblems that are solved via QAOA, with a classical aggregation stage ensuring global feasibility. This structure-aware hybrid design embeds system knowledge into the quantum framework for scalable UC.

In \cite{huangconsensus}, a distributed quantum decomposition is introduced for SCUC with transmission switching. The framework jointly optimizes unit commitments, line switching, and dispatch under N-1 security constraints. Local subproblems are solved classically, while master problems are formulated as QUBOs for quantum solvers. Consensus-based dual updates coordinate regional decisions, enabling efficient hybrid quantum–classical scheduling for SCUC with transmission switching.

In \cite{mahroo2023learning}, a learning-infused quantum–classical optimization framework is proposed for power generation scheduling, combining distributed coordination with quantum-enabled subproblem solving. Local binary subproblems are formulated as QUBOs and solved via parameterized quantum circuits, whose parameters are adaptively tuned by a learner. An ADMM layer enforces inter-area feasibility, yielding a scalable and adaptive hybrid framework for UC-type optimization.

Although \cite{qin2023optimization} does not use quantum hardware, its optimization-assisted ensemble framework mirrors hybrid quantum–classical co-optimization principles. The combination of optimization-guided action generation and multi-agent learning parallels parameter updates in variational quantum circuits, serving as a precursor to quantum-enabled RL systems for future UC scheduling.

In \cite{barrass2025leveraging}, a hybrid quantum–classical algorithm integrates quantum subroutines into a Benders decomposition framework for UC. The master QUBO, solved via a quantum annealer, handles commitment decisions, while classical subproblems refine dispatch and feasibility. Iterative Benders cuts and k-local neighborhood search improve convergence, yielding competitive solutions with reduced computational effort and showcasing the value of quantum integration in UC scheduling.

In \cite{hasanzadeh2025d2}, a hybrid quantum--classical framework is developed for the UC problem using a three-block ADMM that explicitly separates continuous dispatch, binary commitment, and slack consensus updates. The binary block is reformulated as a QUBO compatible with quantum solvers and systematically decomposed into three hierarchical levels---type-specific, micro-, and batched QUBOs---to balance physical coupling and hardware feasibility. The batched formulation aggregates per-unit--time micro-QUBOs into solver-ready groups sized for near-term quantum processors and solves them via a distributed variational quantum eigensolver integrated with an accept-if-better safeguard to stabilize hybrid updates. 

\subsection{QML Approaches: QNN and QRL}
\label{subsec:qml_uc}
Recent advances in quantum machine learning have extended hybrid optimization to data-driven decision-making. These methods use QNNs and quantum reinforcement learning agents to model adaptive UC policies under uncertainty.

In \cite{fu2025quantum}, the quantum-embedded robust UC framework also illustrates the integration of quantum subroutines into learning-based decision processes. Its iterative quantum–classical interaction resembles reinforcement learning, where the quantum circuit generates policies and the classical optimizer supplies feedback and penalty updates. This structure represents a step toward quantum reinforcement learning for resilient UC.

In \cite{zheng2024fast}, the quantum search-based UC formulation also exhibits learning behavior by iteratively refining the probability distribution of feasible commitments through oracle feedback. The oracle acts as a reward signal, guiding the search toward better schedules and linking quantum search with policy learning, thereby paving the way for quantum-enhanced exploration in UC.

In \cite{salgado2024hybrid}, the warm-start mechanism embodies a learning-based hybrid scheme where classical relaxation results iteratively refine the quantum search space. The quantum circuit updates its parameters using classical feedback, a process that resembles reinforcement-style adaptation. This feedback loop bridges hybrid quantum optimization and learning, showing how data-informed initialization enhances convergence in QAOA-based UC solvers.

In \cite{liu2025exact}, an exact quantum UC algorithm is introduced using a PCQNN. Unit commitments are encoded into qubit states processed through structured quantum layers that capture cost and constraint interactions with reduced circuit depth. Parameters are classically optimized to minimize cost and ensure feasibility, showing that partially connected QNNs can achieve near-exact UC schedules with improved circuit efficiency on NISQ hardware.

In \cite{wei2024quantum}, a two-stage UC framework based on QRL is proposed to enhance robustness under renewable and load uncertainty. A quantum deep Q-network handles day-ahead commitments, while a quantum soft actor-critic manages real-time adjustments. Both stages are modeled as Markov decision processes (MDP)s with quantum state encodings and parameterized circuits trained via classical optimization. The QRL framework enhances feasibility and decision speed compared to classical UC methods.

In \cite{mahroo2023learning}, a learning-driven hybrid architecture is presented where classical learning algorithms guide quantum variational solvers. The learning module updates circuit parameters and hyperparameters using feedback from past iterations, embedding adaptive training within the quantum–classical loop. This integration of learning and quantum optimization enhances convergence and robustness in distributed quantum scheduling.

In \cite{qin2023optimization}, an optimization-method-assisted ensemble deep reinforcement learning framework is proposed for large-scale UC under dynamic conditions. The UC is modeled as an MDP, with states capturing the system's status and actions representing commitment decisions. Candidate actions are generated via simplified optimization to guide multiple deep Q-learning agents trained with multi-step updates. The ensemble enhances robustness and generalization, providing a foundation for future quantum reinforcement learning in UC.

\vspace{-3pt}
\subsection{Quantum-Inspired Metaheuristics for UC}
\label{subsec:qmeta_uc}
In parallel with hardware-based methods, several quantum-inspired metaheuristics mimic quantum principles to enhance classical search algorithms for UC.

In \cite{srikanth2018meta}, a quantum binary grey wolf optimizer is proposed for UC. Each candidate is encoded as a qubit vector with probability amplitudes representing on/off states. Quantum rotation gates update these amplitudes while preserving the grey wolf leadership hierarchy. The method minimizes total UC cost with penalty terms for constraints, achieving faster convergence and higher solution diversity without using quantum hardware.

In \cite{ismail2020unit}, a quantum particle swarm optimization algorithm is used for UC as a quantum-inspired metaheuristic. Particles encode commitment schedules via qubit angles, with quantum rotation gates updating amplitudes under a local attractor mechanism guiding convergence. Collapsed binary states are evaluated for cost and constraints, resulting in improved exploration and faster convergence compared to classical PSO.

In \cite{muralikrishnan2020comprehensive}, a comprehensive review of evolutionary algorithms for UC—covering GA, PSO, DE, and GWO—is presented. The survey classifies methods by formulation, constraint handling, and scalability, highlighting how heuristics evolved to tackle nonlinearity and mixed discrete–continuous features. Though focused on classical techniques, it provides a conceptual bridge to quantum-inspired metaheuristics for UC.

In \cite{kawauchi2025development}, a quantum predator–prey brainstorm optimization algorithm is introduced for UC, combining quantum-inspired encoding with predator–prey and brainstorming dynamics. Candidates use qubit-like probabilistic representations updated via quantum rotation operators to enhance diversity. Clustering around elite solutions refines global search, yielding faster convergence and higher robustness than classical heuristics.

In \cite{singh2021towards}, a multipartite adaptive binary and real-coded quantum-inspired evolutionary algorithm is proposed for multi-objective UC with thermal and wind units. Binary qubit-inspired variables encode commitments, while real-coded variables represent dispatch and wind outputs. Quantum rotation operators and adaptive evolution guide Pareto optimization, with a repair mechanism ensuring feasibility.

In \cite{ismail2021solving}, a hybrid PSO framework is proposed for large-scale UC to enhance convergence. Particles encode unit commitments and evolve through adaptive velocity control, mutation-based local search, and constraint-repair mechanisms. The method minimizes total cost under operational constraints, achieving faster convergence and higher solution quality than classic PSO, and serves as a precursor to quantum-inspired UC.

\section{Conclusion and Future Works}
\label{sec:conclusion}
This survey reviewed quantum approaches for the UC problem across four categories: quantum annealing, hybrid variational optimization, quantum machine learning, and quantum-inspired metaheuristics. These works show that quantum methods can accelerate discrete scheduling, reduce search effort, and support data-driven decision-making in UC.

Key strengths observed across the literature include:
\begin{itemize}
    \item UC maps naturally to QUBO and Ising forms, allowing direct use of annealers and variational circuits.
    \item Several studies reduce qubit needs through compact encodings, logarithmic discretization, and fewer variables.
    \item Decomposition methods such as QADMM and QGBD break UC into smaller QUBOs that fit near-term hardware.
    \item Learning-based and warm-started hybrid schemes enhance convergence and better handle uncertainty.
\end{itemize}

Important limitations remain:
\begin{itemize}
    \item Quantum hardware has limited qubits, noise, and restricted connectivity.
    \item UC constraints lead to large and tightly coupled QUBOs that are hard for NISQ devices.
    \item Hybrid methods need careful coordination to maintain feasibility between quantum and classical parts.
    \item QNN and QRL models require long training and lack strong reliability guarantees for large UC cases.
    \item Quantum-inspired methods do not provide true quantum speedup and may struggle with large time horizons.
\end{itemize}

Future research can move toward:
\begin{itemize}
    \item Structure-aware QUBO designs that use system sparsity and constraint patterns to reduce qubit counts.
    \item Distributed quantum computing architectures that divide UC into regional or temporal subproblems executed across multiple quantum processors.
    \item Variational full-quantum optimization algorithms that move beyond hybrid loops and leverage deeper circuits as hardware improves.
    \item Quantum algorithms designed only for UC decomposition.
\end{itemize}

\bibliographystyle{IEEEtran}
\bibliography{example}
	
	
\end{document}